# Crossover Between Weak Antilocalization and Weak Localization and Electron-Electron Interaction in Few-Layer WTe$_2$

Xurui Zhang[1], John M. Woods[2], Judy J. Cha[2], Xiaoyan Shi[1]

[1]*Department of Physics, The University of Texas at Dallas, Richardson, TX, USA*

[2]*Department of Mechanical Engineering and Materials Science, Yale University, New Haven, CT, USA*

We report electron transport studies in an encapsulated few-layer WTe$_2$ at low temperatures and high magnetic fields. The magnetoconductance reveals a temperature-induced crossover between weak antilocalization (WAL) and weak localization (WL) in quantum diffusive regime. We show that the crossover clearly manifests coexistence and competition among several characteristic lengths, including the dephasing length, the spin-flip length, and the mean free path. In addition, low temperature conductance increases logarithmically with the increase of temperature indicating an interplay of electron-electron interaction (EEI) and spin-orbit coupling (SOC). We demonstrate the existences and quantify the strengths of EEI and SOC which are considered to be responsible for gap opening in the quantum spin hall state in WTe$_2$ at the monolayer limit.

Tungsten ditelluride (WTe$_2$), a layered transition metal dichalcogenide (TMD) crystal, has attracted a great deal of interests due to its unique electron transport properties. In the bulk form, WTe$_2$ has been predicted[1] and verified[2] to be the first type-II Weyl semimetal. However, the monolayer form of WTe$_2$ is either a quantum spin Hall insulator[3,4] (QSHI) at low carrier density ($n$) or a superconductor[5] at high $n$. For the QSHI state, a direct band gap emerges in bulk along with the topologically protected edge state. However, it is still unclear how a semimetallic band structure evolves into a gapped structure by reducing the thickness. It is proverbially believed that the band inversion happens in single-layer WTe$_2$, and a band gap is opened due to strong spin-orbit coupling (SOC)[6,7]. However, another study verified that, instead of a full SOC-induced bulk band gap, a Coulomb gap induced by the electron-electron interaction (EEI) was observed[8], which supports the observation of the quantized conduction of topological edge states. In addition, it is well known that EEI becomes stronger as $n$ decreases, thus one can expect EEI should play an important role in the band gap opening. In order to have a better understanding of this mechanism, it is crucial to characterize both the SOC and EEI in WTe$_2$.

In weakly disordered electronic systems, the interaction between different scattering mechanisms gives rise to different transport behaviors. For example, the constructive quantum interference of the electrons moving through time-reversed scattering loops gives rise to a negative quantum correction to the conductance, which is so-called weak localization (WL)[9–11]. On the contrary, the quantum interference induced by SOC gives rise to a positive correction known as the weak anti-localization (WAL)[9–11]. Both WL and WAL can be suppressed by external magnetic field due to the time-reversal symmetry (TRS) breaking. In general, the quantum interference effect and thus its correction on electron conductance weakens as temperature ($T$) increases owing to enhanced decoherence mechanism. However, in some topological insulators, in spite of the existence of WAL effect with positive correction, the conductance decreases as the temperature is lowered[12–17]. This paradox can be solved by taking a strong EEI into account, which is known as the Altshulter–Aronov effect[18]. WTe2 in low $n$ region shows both a strong SOC and a strong EEI, thus this is a good platform to investigate SOC and EEI simultaneously in transport studies.

In this study, we investigate the SOC and EEI in few-layer WTe2 by measuring both the temperature-dependent sheet conductance and magnetoconductance (MC). The strength of SOC can be quantified as the spin-flip length $l_{so}$, the distance travelled by an electron before its spin direction is changed by the scattering[19,20]. The EEI can be characterized by the Coulomb screening factor $F$ which is a measure of the screened Coulomb interaction[21,22]. We also observe a temperature-induced crossover between WAL and WL. The mechanism of the crossover is clearly interpreted based on the relative change between spin-flip length $l_{so}$ and dephasing length $l_{\phi}$, the distance travelled by an electron within which it can maintain the phase coherence.

The thin WTe2 flake, which was about 5 nm thick, was obtained by mechanical exfoliation of bulk WTe2 crystals synthesized by chemical vapor transport[23]. The exfoliated sample was encapsulated between two pieces of hexagonal Boron Nitride (hBN) thin flakes which were about 10 nm thick and transferred onto a silicon substrate with 285 nm SiO2 coating on surface by dry transfer technique[24]. Ultrathin WTe2 flakes are very easy to get oxidized in air[5,25]. Hence the hBN flakes are necessary here to protect the sample from air. In addition, they provide a cleaner interface for WTe₂. The electron-beam lithography was used to make a pattern and the ohmic contacts were deposited by electron-beam evaporation of Pd/Au (10 nm/50 nm) followed by a lift-off process. Transport measurements down to $0.036\,\text{K}$ were carried out in an Oxford dilution refrigerator. Both the longitudinal resistance $R_{xx}$ and Hall resistance $R_{xy}$ were measured simultaneously by using standard low frequency lock-in techniques. The corresponding four-probe measurement setup is shown in the inset of figure 1(a).

We first investigate the temperature dependence of sheet conductance, as shown in figure 1(a), at zero magnetic field ($H = 0\,\text{T}$). We find two distinct regions divided by a critical temperature around $T_{\max} \approx 13.8\,\text{K}$ as shown in the inset of figure 1(a). Above this temperature, the resistance exhibits a typical metallic temperature dependence, which agrees with the expected semi-metal properties of bulk WTe₂. However, an up-turn occurs at $T_{\max}$, which indicates a tendency of weak localization. Fitting result shows that the low-temperature conductance is proportional to $\ln(T/T_{\max})$. Such relation has been observed in some topological materials[12–17]. In those experiments, a suppression of the conductance with decreasing temperature, like the case in this device, is observed along with weak antilocalization effect observed in magnetoconductance measurements which would enhance the conductance. Such a seeming paradox in topological insulator can be interpreted by the interplay of disorder and electron-electron interaction, which is known as the Altshulter–Aronov effect[18]. The correction from electron-electron interaction to the conductance would decrease logarithmically with decreasing temperature. The conductance enhancement from weak antilocalization could be overwhelmed which leads to an overall weak localization tendency. We further investigate the temperature dependent sheet conductance at various magnetic field as shown in figure 1(b). We find that the conductance at low temperature can be well fitted by equation $\sigma \sim \kappa \ln T$, where fitting parameter $\kappa$ raises and saturates quickly as magnetic field increases (figure 1(b) inset). A detailed analysis will be provided in the latter part.

We also carry out magnetoconductance measurements to further investigate the weak antilocalization effect in perpendicular magnetic field. The magnetoconductance (MC) is defined as $\Delta\sigma(B) = \sigma(B) - \sigma(0)$, where $B = \mu_0 H$ is the effective magnetic field, at various constant temperatures, shown in figure 2(a). At high magnetic field, the sample exhibits quasi-parabolic MC which is common in WTe$_2$[25–30]. However, at low magnetic field, the parabolic MC doesn't dominate any more. At 0.036 K, a negative cuspate MC shows up at low field, which is a typical characteristic of WAL effect[9,11]. With temperature increasing, the cusp in MC gradually broadens until 11 K, beyond which the MC becomes positive first then decreases again with the magnetic field increasing. The positive MC corresponds to the WL effect[9,11]. This suggests that not only is there a temperature-dependent crossover between WL and WAL but it is very likely that WL and WAL coexist and compete. In order to better understand the characteristics of the sample, we carry out Hall measurements at different temperatures. The Hall resistance ($R_{xy}$) curves at different constant temperatures are shown in figure 2(b). These tight overlapping curves indicate that the carrier density is insensitive to temperature. The inset shows the $R_{xy}$ curve at 0.036 K and the corresponding linear fit, which gives the sheet charge density $n = 6.25 \times 10^{13}$ cm$^{-2}$ and carrier mobility $\mu = 58.7$ cm$^2$/Vs. Therefore, this device is in a low $n$ region and the charge transport is dominated by quantum diffusions. Furthermore, the mean free path is $l = \hbar k_F \mu / e = 7.6$ nm, where $\hbar$ is the reduced Plank constant, $k_F$ is the Fermi wave vector, and $e$ is the electron charge.

To analyze the temperature-dependent competition between WL and WAL, we introduce the two-component Hikami-Larkin-Nagaoka (HLN) theory for a 2D system[9,10,31]

$$\Delta\sigma(B) = \sum_{i=0,1} \frac{\alpha_i e^2}{\pi h} \left[ \Psi\left(\frac{l_B^2}{l_{\phi_i}^2} + \frac{l_B^2}{l_i^2} + \frac{1}{2}\right) - \ln\left(\frac{l_B^2}{l_{\phi_i}^2} + \frac{l_B^2}{l_i^2}\right) \right], \qquad (1)$$

where $\Psi$ is the digamma function, $\alpha_0$ and $\alpha_1$ stand for the weights of WL and WAL respectively. In the limit of pure WAL, $\alpha_0 = 0$ and $\alpha_1 = -1/2$ for each band carrying a $\pi$ Berry phase[9,11,31–35], while in the limit of pure WL, $\alpha_0 = 1$ and $\alpha_1 = 0$ for a usual 2D system and $\alpha_0 = 1/2$ and $\alpha_1 = 0$ for a topological surface state[9,11,31–35]. In the case of a coexistence of WL and WAL, $\alpha_0$ and $\alpha_1$ could take intermediate values. $l_B = \sqrt{\hbar/4eB}$ is the magnetic length and all other characteristic lengths take the same form. $l_{\phi_i}$ is the corresponding dephasing length and $l_i$ gives correction to $l_{\phi_i}$. In WAL where SOC plays an important role, $l_i = l_{so}$, while in WL where SOC has nothing to do with, the $l_i$ can be neglected. $l_{so}$ is the spin-flip length, which describes the strength of SOC[19,20,36]. Of course there are some other mechanisms that can be included in $l_i$, like magnetic scattering which needs to be discussed in some magnetically doped samples[34,37–41], but this does not apply to our situation. In addition, considering the particularity of WTe$_2$ in which the intrinsic quasi-parabolic positive magnetoresistance (MR) originating from carrier compensation cannot be ignored, we introduce a quadratic term as the background[25–30], such that

$$\Delta\sigma(B)_{FIT} = \Delta\sigma(B) + cB^2 \qquad (2)$$

where $c$ is a constant that depends on the measurement temperature. To all the temperature-dependent $\Delta\sigma(B)$ traces, Eq. (2) fits the data in low-field region (-2 T to 2 T) very well. For clarity, the fits are shown in Fig. 3(a) and (b) for WAL-dominant and WL-dominant regimes, respectively. The coincidence of the fitting curves and the experimental data validates the perfect applicability of the HLN model.

Figure 4(a) shows the evolution of $|\alpha_0|$ ($\alpha_0 > 0$) and $|\alpha_1|$ ($\alpha_1 < 0$), obtained from the two-component HLN fitting, as a function of temperature. From 0.036 K to 25 K, $\alpha_1$ changes slowly from -0.5 to -0.44 while $\alpha_0$ increases drastically from 0.16 to 0.5. The half integer values of $\alpha_0$ and $\alpha_1$ indicate the existence of the topological surface state in WTe$_2$ ultrathin film. It's worth mentioning that the very small change in $\alpha_1$ and dramatic change in $\alpha_0$ indicate that WL is much more sensitive to the temperature than WAL.

In order to investigate the mechanism of the crossover and competition between WL and WAL, we further examine the changes in several characteristic lengths with temperature. Temperature dependence of the dephasing lengths for WL ($l_{\phi_0}$) and WAL ($l_{\phi_1}$) and the spin-flip length ($l_{so}$) are shown in Fig. 4(b). The rapid reduction of the dephasing lengths indicates that the inelastic scattering in this sample is enhanced drastically as $T$ increases. It is also of note that the dephasing lengths are much larger than the film thickness and the mean free path ($l = 7.6 \text{ nm}$) even at high temperature, which confirms the transport measurements are indeed in the 2D quantum diffusion regime. Power-law fits to the dephasing lengths in logarithmic coordinate (inset of Fig. 4(b)) give $l_{\phi_0} \sim T^{-0.25}$ for WL and $l_{\phi_1} \sim T^{-0.23}$ for WAL. The almost identical temperature dependence is expected since the electrons participating in WL and WAL pass through the same or similar TRS loops and undergo the same inelastic scattering which depends on the temperature only. The only difference is that the electrons participating in WAL undergo frequent spin flips which generate a $\pi$ Berry phase after moving through a scattering loop. This $\pi$ Berry phase is responsible for the destructive quantum interference that suppresses the back scattering and enhances the conductance, leading to the WAL. In present case, the effective exponent of temperature p is ~ 0.5 in $l_\phi \sim T^{-p/2}$ which is considerably lower than that theoretically expected $p$ exponent ($p = 1$) for the Nyquist electron-electron scattering process in 2D[42,43]. However, a similar p ~ 0.5 has been observed in topological insulator Bi$_2$Se$_3$ micro-flakes, where EEI plays important roles[14,44,45]. Based on the temperature dependence of conductance and the power law dependence of $l_\phi$, we believe that there exist additional electron dephasing processes which are noticeable over the experimental temperature range in our sample as well.

Now we focus on the mechanism of the crossover between WL and WAL. Since the $\pi$ Berry phase induced by SOC is the key to determining whether a sample presents WL or WAL, we plot $l_{so}$, which reflects the strength of SOC and the band topology, as a function of temperature in Fig. 4(b). We find that $l_{so}$ can be well described with power-law fit ($l_{so} \sim T^{-0.14}$) below a characteristic temperature $T_{so} \sim 11 \text{ K}$ (Fig. 4(b) inset), which is exactly the WAL dominant regime. Beyond $T_{so}$, the $l_{so}$ decreases even faster and deviates from the power-law decay as temperature increases. In contrast, both $l_{\phi_0}$ and $l_{\phi_1}$ maintain the power-law decay up to 25 K, which indicates the quantum interference can survive in a much higher $T$ than SOC. In high $T$, where $l_{so}$ approaches to and even smaller than the mean free path, the symmetry class changes from symplectic to orthogonal[46,10] and thus the quantum interference correction will crossover from WAL to WL as temperature increases. In a topological insulating system, the effect of quantum interference can be characterized by two time scales[36]: the dephasing time $\tau_\phi$ and the spin-flip time $\tau_{so}$ with the relation $l_j^2 = D\tau_j$ where $l_j$ can be $l_\phi$ or $l_{so}$, and $D$ is the electron diffusion coefficient. Let's evaluate the characteristic lengths and related time qualitatively in both WL and WAL processes. In the regime $\tau_\phi \gg \tau_{so}$, the spin of the electron undergoes very frequent flips which makes destructive quantum interference between the TRS scattering loops, leading to WAL. In the regime where $\tau_\phi$ is comparable to $\tau_{so}$ or $\tau_\phi < \tau_{so}$, the spin orientation is not that frequently changed by the scattering. In such case, constructive quantum interference occurs as a WL feature. In the intermediate regime where $\tau_\phi > \tau_{so}$, there will be a situation where WL and WAL may coexist. since $\tau_\phi$ and $\tau_{so}$ could vary with temperature, this variation signifies a crossover between WAL and WL. In present sample, we have $l_\phi \approx 175$ nm (taken from $l_{\phi_1}$ in WAL dominant regime), $l_{so} \approx 24$ nm $\gg l$ at 0.036 K and $l_\phi \approx 30$ nm (taken from $l_{\phi_0}$ in WL dominant regime), $l_{so} \approx 7$ nm $\approx l$ at 25 K. These fitting results give $\tau_\phi/\tau_{so} \approx 53$ at 0.036 K and $\tau_\phi/\tau_{so} \approx 18$ at 25 K. We conclude that our sample is in an ambiguous state between WAL regime in which $\tau_\phi \gg \tau_{so}$ and intermediate regime where $\tau_\phi > \tau_{so}$ at 0.036 K. With the temperature increasing, the difference between $\tau_\phi$ and $\tau_{so}$ is narrowing and our sample enters into the intermediate regime, giving rise to the co-existence of WL and WAL. This sample never goes into pure WL regime which requires $\tau_\phi < \tau_{so}$ and presents a sharply downward cuspate MC.

In previous discussion of MC, we don't take EEI into consideration. Although the role of EEI dominates in temperature dependence of conductance, MC due to EEI is at least one order smaller than that due to quantum interference in perpendicular magnetic field[11,13,16,17]. The EEI affects MC indirectly via the Zeeman splitting. The correction to the conductance is $\Delta\sigma_{EEI}(B) = \frac{e^2}{\pi h}\frac{\tilde{F}^\sigma}{2}g_2\left(\frac{E_Z}{k_B T}\right)$, where $\tilde{F}^\sigma$ is the Coulomb screening factor, $g_2$ is an integral function, and $E_Z$ is the Zeeman energy[13,17]. In the regime where EEI are important, the Zeeman contribution could be strongly suppressed by SOC. Theories[47,48] and experiments[49] on other materials have unambiguously shown that strong SOC can diminish and even entirely suppress the Zeeman-split term in the diffusion channel[13].

In contrast, in analysis of temperature-dependent sheet conductance, it's necessary to consider both EEI and quantum interference. The quantum correction to the conductance resulting from WAL at zero magnetic field is given as[12,50,51]

$$\Delta\sigma_{QI}(T) = \alpha p \frac{e^2}{\pi h}\ln\left(\frac{T}{T_{QI}}\right) \qquad (3)$$

where $\alpha$ is exactly the same parameter discussed in HLN model and $p$ is the exponent in the power-law fit of dephasing lengths. $T_{QI}$ is a characteristic temperature at which the quantum correction vanishes. The correction to the conductance coming from EEI is given by[11,12,14,16,17]

$$\Delta\sigma_{EEI}(T) = \frac{e^2}{\pi h}(1-\frac{3}{4}F)\ln\left(\frac{T}{T_{EEI}}\right) \qquad (4)$$

where $\tilde{F}$ is the simplified Coulomb screening factor, which is not critical to be distinguished with $\tilde{F}^\sigma$ in most experiments[21,22]. $T_{EEI}$ is the characteristic temperature for the EEI effect. In most experiments, $T_{QI}$ and $T_{EEI}$ are considered to be the same and to be the turning point in the temperature-dependent conductance, which is $T_{max} = 13.8\text{ K}$ in our case. Having summed the WAL and EEI contributions together, we got the slope to be $\alpha p + (1-\frac{3}{4}F) = 0.53$ at 0 T, obtained from the fitting $\Delta\sigma(T) \propto \ln(T/T_{max})$ shown as the black dashed line in Fig 1(a). Since the sample is in WAL dominant regime below 13.8 K, we only take the weight of WAL, $\alpha_1$, which is almost unchanged between -0.5 and -0.47. Having $p = 0.46$ obtained from the aforementioned fit, we got the Coulomb screening factor $0.32 \leq F \leq 0.34$. This F value is in good agreement with the theory that F is between -1 and 1 with $F \rightarrow 1$ in the limit of complete screening (good metal) and $F \rightarrow 0$ in the limit of no screening (bad conductor)[22]. The fitting slopes at various magnetic fields shown in the inset of figure 1(b) agree well with Chen's results[13], which is theoretically verified to be a single topological surface channel[11]. Since the quantum interference is strongly suppressed at high magnetic field, the slope at 12 T should be totally attributed to EEI. By only fitting with equation (4), we get the Coulomb screening factor F to be 0.32 which agrees with the estimated value at 0 T. This also confirms our analysis of MC. In some topological materials, people got negative F, which might be related to strong electron-phonon coupling[21,52] or contribution from bulk state[12]. Even though strong SOC is confirmed in this sample, we still get a reasonable value of F in the expected range which is in good agreement with theory. This agreement strongly indicates the very first observation of electron-electron interaction in few-layer WTe$_2$ in transport studies. Similar existence of EEI has been confirmed in monolayer samples[8]. In monolayer WTe$_2$, the opening of the Coulomb gap induced by EEI can diminish the bulk state and support the quantized conduction of topological edge states[8]. Unlike the monolayer samples, this few-layer WTe$_2$ device shows both a semi-metallic $\sigma(T)$ at high T as expected for bulk and an EEI induced $\ln(T)$ behavior at low T as expected for monolayer. This temperature-tuned crossover in the conductance bridges the conductance in two extreme cases and provides a uniform picture of understanding in the transport behavior in WTe$_2$.

In conclusion, we have investigated the quantum interference and electron-electron interaction in encapsulated few-layer WTe$_2$. At low temperatures, a clear WAL to WL crossover revealed a manifestation of coexistence and competition among several characteristic lengths. With increasing either temperature or magnetic field, the topologic non-trivial transport fades out and a related Coulomb gap closes. In this process, both SOC and EEI play important roles. In addition, quantum interference, and thus TRS, can survive in a much higher temperature, which makes WTe$_2$ a promising platform for further investigation in the interplay of SOC and band topology. As an intermediate form between the monolayer and the bulk limits, few-layer devices could behave as either limit by simple tuning of temperatures. This unique feature may enable new spintronic applications.

Acknowledgement

This work was supported by UT Dallas research enhancement fund. Synthesis of $WTe_2$ crystals was supported by DOE BES Award No. DE-SC0014476.

dimensional limit. *Phys. Rev. B* **83**, 165440 (2011).

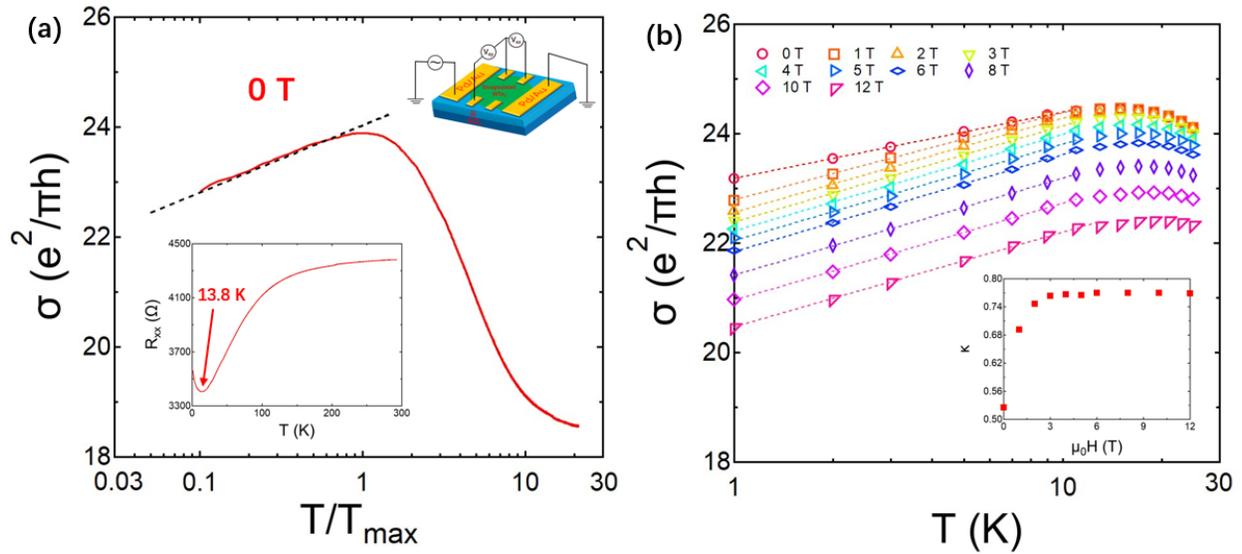

Figure 1. (a) Temperature dependence of sheet conductance at zero magnetic field (red solid line). The black dashed line is the logarithmic fit at low temperature regime (T<13.8 K). Insets show the temperature-dependent resistance (lower left) and a schematic diagram of the measurement setup (upper right). The red arrow marks the up-turn at 13.8 K. (b) Temperature dependence of sheet conductance (hallow symbols) at various magnetic fields. The dashed lines are the logarithmic fits at low temperature regime. Inset shows the slopes $\kappa$ as a function of magnetic field.

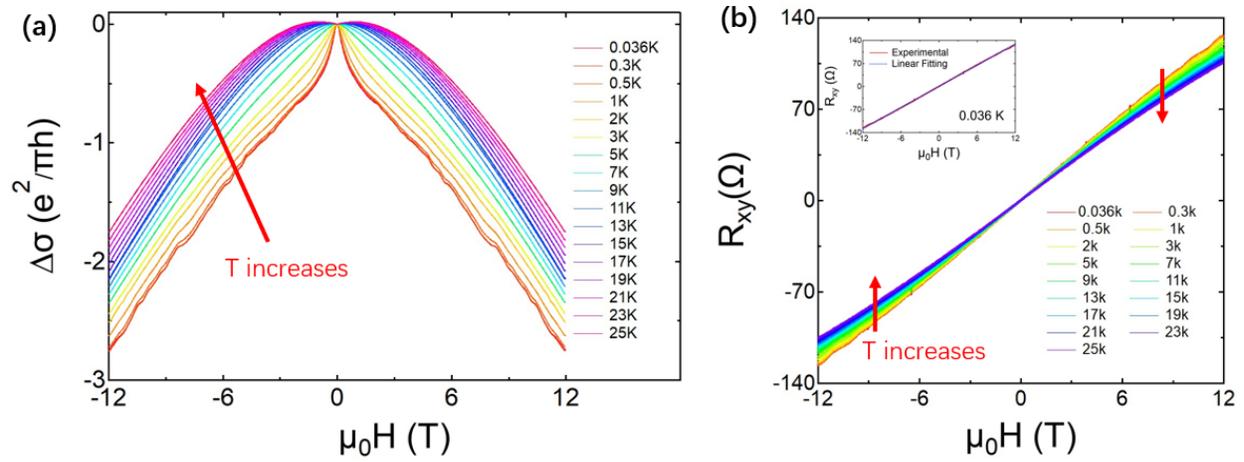

Figure 2. (a) Temperature dependence of the magnetoconductance, defined as $\Delta\sigma(H) = \sigma(H) - \sigma(0)$, in unit of $e^2/\pi h$. (b) Temperature dependence of the Hall resistance. Inset is the linear fit to the Hall resistance at 0.036 K which gives $n = 6.25 \times 10^{13}$ cm$^{-2}$. Red arrows in both panels denote the direction of increasing temperature.

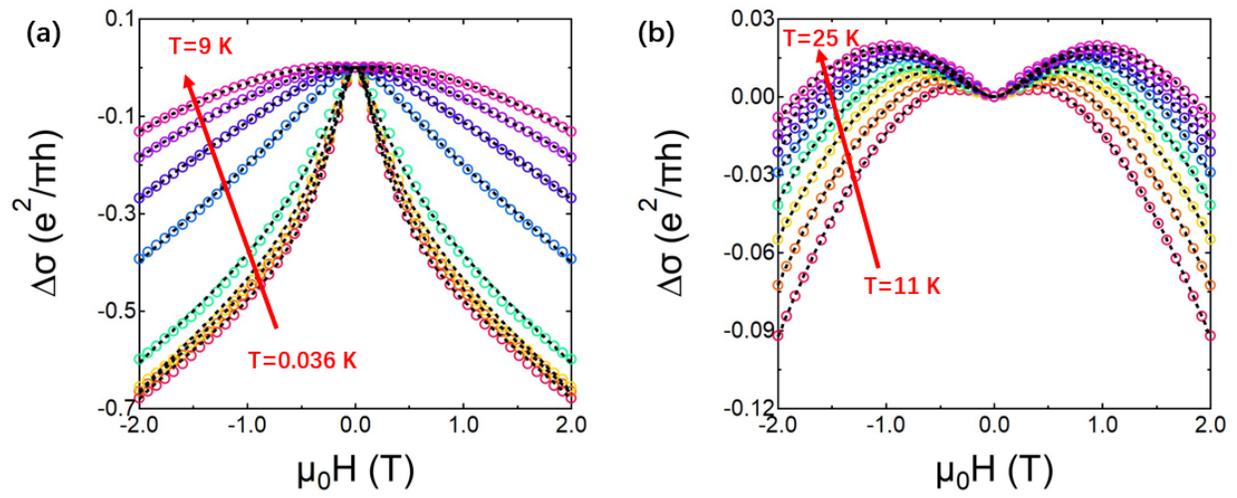

Figure 3. Magnetoconductance $\Delta\sigma(H)$ at (a) low temperatures (0.036 K to 9 K) and (b) high temperatures (11 K to 25 K). Hollow circles and dashed lines are experimental data and two-component HLN fits, respectively. Red arrows indicate the direction of increasing temperature.

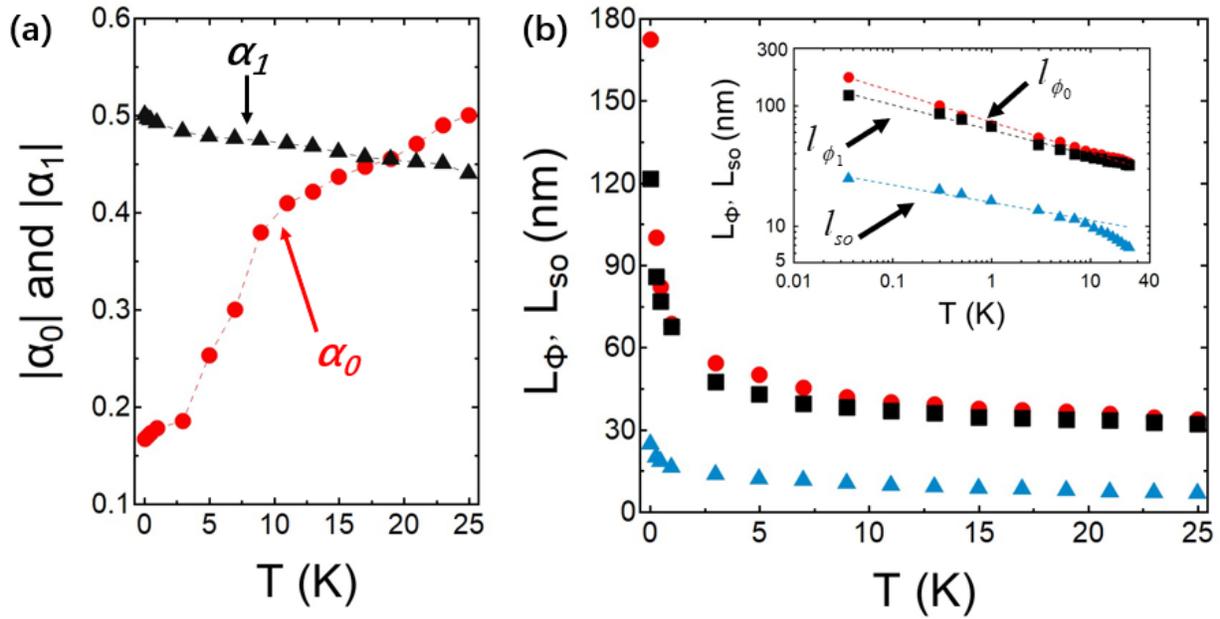

Figure 4. Temperature dependence of (a) $|\alpha_0|$ and $|\alpha_1|$, and (b) $l_{\phi_i}$ and $l_{so}$. Inset of (b): the same set of data, as in the main plot, are shown in log-log plot. Dashed lines are the power-law fits.